\documentclass[%
 aip,
 amsmath,amssymb,
reprint,%
]{revtex4-1}

\usepackage{graphicx}
\usepackage{epsfig}
\usepackage{url}

\usepackage{graphicx}
\usepackage{dcolumn}
\usepackage{bm}

\usepackage[utf8]{inputenc}
\usepackage[T1]{fontenc}
\usepackage{mathptmx}

\begin{document}


\title{Indexing current-voltage characteristics using a hash function}

\author{T. Tanamoto}
\affiliation{Department of information and science, Teikyo University, 
1-1 Toyosatodai, Utsunomiya, Tochigi 320-8551, Japan} 
 \email{tanamoto@ics.teikyo-u.ac.jp}
\author{S. Furukawa}
\affiliation{Department of information and science, Teikyo University, 
1-1 Toyosatodai, Utsunomiya, Tochigi 320-8551, Japan} 

\author{R. Kitahara}
\affiliation{Department of information and science, Teikyo University, 
1-1 Toyosatodai, Utsunomiya, Tochigi 320-8551, Japan} 

\author{T. Mizutani}
\affiliation{Institute of Industrial Science and the VLSI Design and Education Center, University of Tokyo, Tokyo, Japan} 

\author{K. Ono}
\affiliation{Advanced Device Laboratory, RIKEN, Wako-shi, Saitama 351-0198, Japan}

\author{T. Hiramoto}
\affiliation{Institute of Industrial Science and the VLSI Design and Education Center, University of Tokyo, Tokyo, Japan}

\date{\today}

\begin{abstract}
Differentiating between devices of the same size is essential for ensuring their reliability. 
However, identifying subtle differences can be challenging, particularly when the devices share similar characteristics, such as transistors on a wafer. 
To address this issue, we propose an indexing method for current-voltage characteristics that assigns proximity numbers to similar devices. 
Specifically, we demonstrate the application of the locality-sensitive hashing (LSH) algorithm to Coulomb blockade phenomena observed
in PMOSFETs and nanowire transistors. 
In this approach, lengthy data on current characteristics are replaced with hashed IDs, facilitating identification of individual devices, and streamlining the management of a large number of devices.
\end{abstract}

\maketitle

As transistor sizes continue to shrink, the number of transistors that factory engineers must analyze increases significantly~\cite{Moore}. Classifying devices within a wafer is essential, 
as their characteristics can vary depending on their location, such as whether they are positioned at the edge or center of the wafer~\cite{Schram,Quach,Nam}. 
As data volume grows, managing this vast amount of information becomes increasingly complex. 
While automated measurements combined with statistical data analysis are now standard practices~\cite{Kim,Calik,Fang}, 
the detailed analysis of individual devices still heavily depends on the expertise of experimenters and engineers.


A Physically Unclonable Function (PUF) is an important hardware security mechanism that protects personal information by providing a unique ID for each device at a low cost~\cite{Guajardo,Holcomb,Marukame,Suh,wPUF,Chen}. 
Smaller devices are more susceptible to defects and traps, leading to greater variability among them. Controlling atomic order is either impractical or prohibitively expensive~\cite{Nagano}.
In Ref.~ \cite{tanaPUF}, we proposed a PUF that utilizes the effects of trap sites, which can be observed in the subthreshold region as a fingerprint of transistors. 
Distinguishing devices using image recognition algorithms has the advantage of enabling the direct application of a wide range of recognition techniques, including artificial intelligence.
In Ref.~ \cite{tanaPUF}, we employed {\it OpenCV} algorithms~\cite{OpenCV, AKAZE, BRISK, ORB} 
to differentiate between chips based on variations in the Coulomb diamond images. However, this method requires storing image data, which generally demands a large amount of memory. 
Therefore, developing a more efficient method for distinguishing devices is crucial for applying PUF to nanodevices.

To efficiently manage a large number of devices including the PUF application, assigning each device a unique identifier based on its characteristics would be beneficial. 
A direct approach is to apply clustering techniques to the current-voltage ($I$-$V$) characteristics, enabling the visualization of differences in device performance~\cite{python}. 
However, conventional clustering may not always effectively distinguish subtle differences between devices (see appendix). Instead of analyzing all $I$-$V$ characteristics individually, 
assigning an ID that encapsulates the current characteristics of each device would simplify management and reduce the engineers' workload.

Locality-Sensitive Hashing (LSH) is a method designed for searching data in high-dimensional spaces by increasing the likelihood that similar data points share the same hash value\cite{Indyk1,Indyk2,Datar2004,Lu,Tian}. 
Indyk and Motwani introduced this technique to mitigate the "curse of dimensionality"~\cite{Indyk1}. 
The effectiveness of LSH has been demonstrated in various applications, including image search\cite{Zhang2016}, document similarity search, genetic data analysis~\cite{Cai}, music recommendation systems~\cite{Luo}, 
and anomaly detection in sensor data~\cite{Charyyev}. 
For example, image search engines utilize LSH to rapidly identify images similar to those uploaded by users within extensive databases. 
LSH filters have been proposed for video anomaly detection~\cite{Zhang2016} and have also been applied to detect anomalies in sensor data collected from IoT devices~\cite{Charyyev}.

In this study, we apply LSH to device datasets and investigate the appropriate hash function parameters that best fit the experimental data. 
We evaluate two distinct types of devices: PMOSFETs operating within the Coulomb blockade regions and nanowire transistors. 
Coulomb blockade is a significant phenomenon observed in nanodevices, and the Coulomb diamond patterns reveal clear differences between devices, allowing for precise identification and classification. 
By contrast, nanowire transistors exhibit monotonous $I$-$V$ characteristics, making differentiation more challenging. We demonstrate that LSH effectively distinguishes these devices. 
The characteristics of devices can shift owing to various factors, and measurements taken on different days may yield different results for the same device. 
This variability is particularly relevant for PUF applications. We compare the hash IDs derived from the $I$-$V$ characteristics, highlighting the robustness and reliability of our approach.

\begin{figure}
\centering
\includegraphics[width=8.5cm]{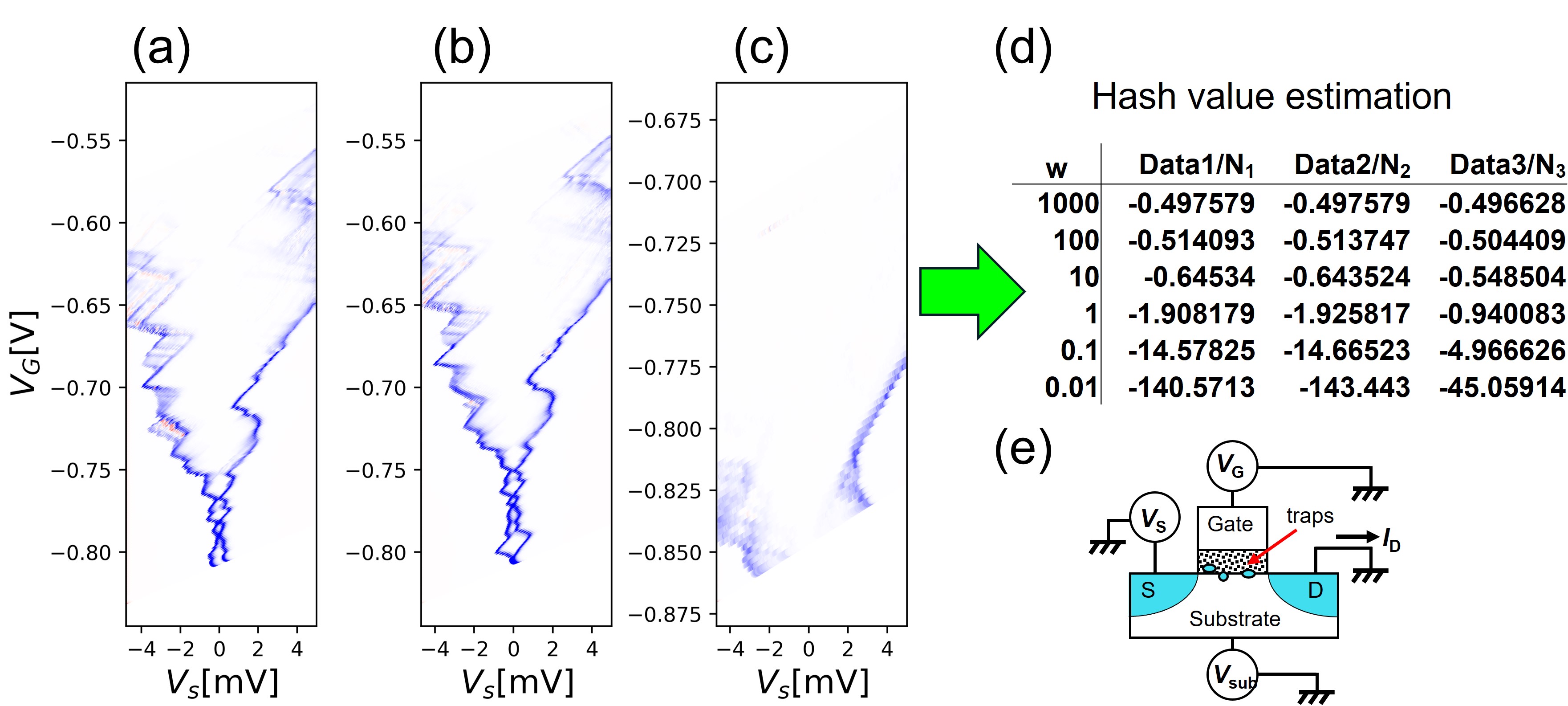}
\caption{
(a)(b)(c)
Coulomb diamonds were observed in the differential conductance characteristics \((dI_D/dV_S)\) of trap states in a conventional pMOSFET 
with a channel length \(L = 125 \, \text{nm}\) and width \(W = 220 \, \text{nm}\) at a temperature of \(T = 1.54 \, \text{K}\). 
A silicon oxynitride layer was used for the gate dielectric. 
Although these devices share the same layout parameters (length and width), 
they exhibit different characteristics owing to variations in the distribution of trap sites. 
Additionally, the hash values for datasets (a), (b), and (c) are provided in section (d). 
\(N_i\) represents the number of data points for each dataset.
}
\label{figCB}
\end{figure}

{\it Hash function}---
In LSH, a given data is converted by the hash function defined by~\cite{Indyk1}
\begin{equation}
h_{{\bf a},b}({\bf v})=\left\lfloor \frac{{\bf a} \cdot {\bf v}-bw}{w} \right\rfloor
\label{hash}
\end{equation}
where ${\bf v}$ represents the given dataset, which in our case consists of the experimentally obtained $I_{\rm D}$ values. 
The parameters ${\bf a}$ and $b$ are a randomly generated vector and a random number within the range $\{0,1\}$, respectively. 
The segment size, $w$ , determines how the given data is partitioned. Additionally, multiple sets of ${\bf a}$ and $b$ are prepared, which are called "bucket", 
and are given by 
$g({\bf v})=\{ h_{{\bf a}_1,b_1}({\bf v}),h_{{\bf a}_2,b_2}({\bf v}),...,h_{{\bf a}_{n_{buk}},b_{n_{buk}}}({\bf v})\}$, 
where $n_{buk}$ is the number of elements in the bucket.
Conventionally, many sets of the buckets are prepared, such as $\{ g_0({\bf v}), g_1({\bf v}),...\}$; 
however, here we only consider one $g({\bf v})$ to observe what kinds of parameters are appropriate 
for distinguishing the devices.

Conventional datasets, such as images, are typically represented using binary values {0, 1}, 
and the distance between data points is computed based on these binary representations. By contrast, $I$-$V$ characteristics are expressed as real numbers, often in exponential notation. 
Although one approach is to discretize these real numbers into digital values, this introduces additional computational overhead and potential loss of precision. 
Therefore, it is preferable to process the raw experimental data directly to maintain accuracy and minimize unnecessary transformations.

In Eq.~(\ref{hash}), $w$ represents the width of the bucket, which plays a crucial role in distinguishing similar data points while ensuring that distinct data points are assigned to separate buckets. 
Selecting an appropriate value for $w$ is essential for effectively differentiating between device data. 
In the following section, we will determine suitable values for $w$ by first estimating the direct Euclidean distance among the data. 
Our analysis will ultimately show that the optimal choice for $w$ is the standard deviation of the raw data.

One of the key advantages of using LSH is its ability to significantly reduce the number of data points that require processing. 
Given $N$ devices and defining $n_d$ as the number of elements in the $I$-$V$ dataset, conventional processing would involve handling $Nn_d$ data points. 
However, by assigning hash IDs to each $I$-$V$ characteristic, the computational burden is drastically reduced to just $N$ numerical values, making the approach highly efficient for large-scale applications.

{\it Results}---
We show the result of the application of LSH to Coulomb blockade and nanowire transistors, 
and the appropriate choice of the hash parameter $w$ is estimated.

{\it Coulomb blockade}---
Figure 1 illustrates the Coulomb diamond structure that arises owing to trap sites at the Si/SiO${}_2$ interface in a 130 nm PMOSFET~\cite {tanaPUF,Ono1,Ono2}. 
Figures 1(a) and 1(b) display Coulomb diamonds measured from the same device on different days, while Figure 1(c) presents data from a different device. 
Despite having identical gate lengths and widths and being fabricated on the same wafer, the devices exhibit distinct Coulomb diamond characteristics due to variations in trap profiles. 
These uncontrollable trap distributions provide a unique fingerprint, making them well-suited for PUF applications. 
Additionally, the discrepancy between Figures 1(a) and 1(b) is attributed to changes in the measurement environment, 
such as variations in the applied force from the needle of the apparatus or fluctuations in temperature.

First, we estimate the Euclidean distance between the datasets, 
where the Euclidean distance between the two data points, \( A \) and \( B \), is defined by the equation:
\begin{equation}
d_{A,B} = \sqrt{ \sum_{j=1}^{n_d} \left( \frac{dI_{D,j}^A}{dV_G} - \frac{I_{D,j}^B}{dV_G} \right)^2}.
\label{d1}
\end{equation}
The dataset shown in Fig. 1 is represented as \( (V_{\rm G}, V_{\rm D}, \frac{dI_{\rm D}}{dV_{\rm G}}) \).
To facilitate analysis, this data is converted into a two-dimensional representation 
with coordinates \( (x,y) \), 
where \( y \) corresponds to \( \frac{dI_{\rm D}}{dV_{\rm G}} \).
Then, the pair \( (V_{G}, V_{D}) \) is treated as a single axis given by \( x = (V_{G}, V_{D}) \). 
The number \( n_d \) refers to the total data points, 
calculated as the product of the number of \( V_{\rm G} \) values and the number of \( V_{\rm D} \) values. 
To ensure a fair comparison between the two datasets, we must use a consistent \( x \)-axis. 
However, the current data for \( x = (V_{G}, V_{D}) \) shown in Figs. 1(a) to 1(c) do not align. 
To address this issue, we identify and select a common \( (V_{G}, V_{D}) \) region that overlaps across the three datasets.

The Euclidean distances between the data points are calculated by \( d_{(a),(b)} = 1012.78 \), \( d_{(b),(c)} = 1061.95 \), and \( d_{(a),(c)} = 1297.23 \). 
The difference in the Euclidean distance between the same devices ((a) and (b)) is approximately 4.74\%, 
whereas the difference between the different devices ((a) and (c)) is approximately 25.1\%. 
This demonstrates that the Euclidian distances serve as a metric to distinguish between devices.
However, as the number of devices \( N \) increases, the Euclidean distances \( {}_N C_2 = \frac{N(N-1)}{2} \) must be managed, 
which is inefficient. Therefore, indexing devices is crucial. 

To address this issue, we apply Eq.(\ref{hash}) to the three datasets shown in Fig. 1. The hash function in Eq.(\ref{hash}) includes a parameter \( w \) that needs to be determined. 
Figure 1(d) presents the hash values for the three Coulomb diamonds using various choices of \( w \). The results indicate that smaller values of \( w \) are more effective in differentiating between the devices. 
Notably, a Euclidean distance of approximately 1000 is insufficient to distinguish between datasets (a) and (c), necessitating an alternative approach.
The standard deviations of the original data are estimated as follows: \( std_{(a)} = 8.463 \), \( std_{(b)} = 3.087 \), and \( std_{(c)} = 8.916 \). 
This suggests that selecting \( w \) based on the standard deviation, rather than the absolute Euclidean distance, is more appropriate. 
From these observations, we conclude that:(1). The same device produced similar hash values.
(2). Different devices yield distinct hash values when \( w \) is in the order of the standard deviation.


\begin{figure}
\centering
\includegraphics[width=8.8cm]{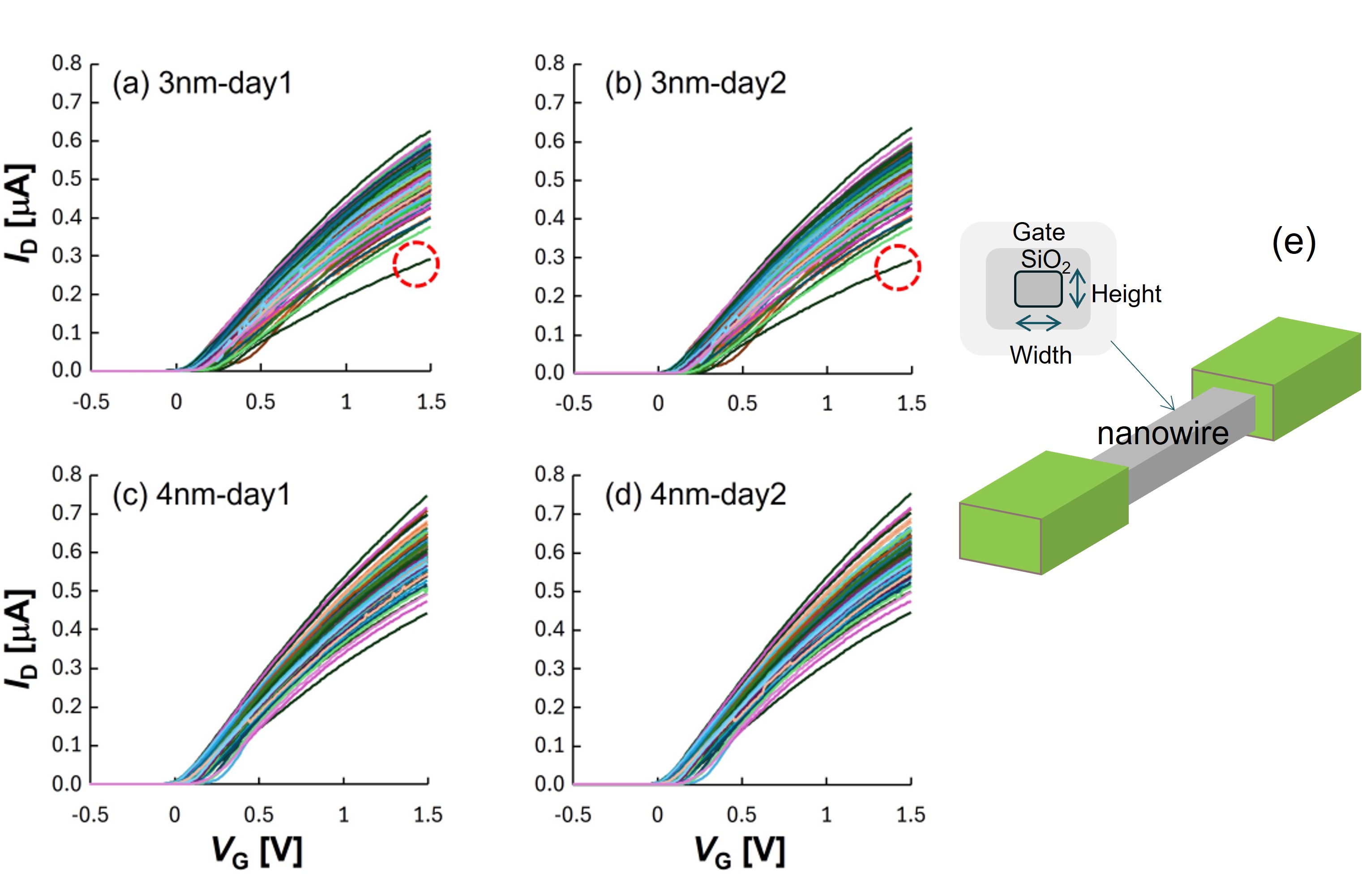}
\caption{
$I_{\rm D}$-$V_{\rm G}$ characteristics of the GAA silicon nanowire transistor at $V_{\rm D}=50$mV.
(a)(b) 3 nm width, and (c)(d) 4 nm width.
No.141 (marked by a red circle) in both (a)(b) deviate from others.
(e) Schematic of GAA silicon nanowire FET,
where $L$=100 nm, $H$=3 nm.
}
\label{fignanowire}
\end{figure}

{\it Nanowire transistors}---
The nanowire transistors do not exhibit specific characteristics, such as Coulomb diamonds. 
Gate-all-around (GAA) silicon nanowire nFETs with (110)-oriented channels were fabricated on (001) SOI substrates. 
The height of the nanowire channel ($H$) is 3 nm, and the channel length ($L$) is 100 nm~\cite{hiramoto1}. 
The detailed fabrication process is described in Refs. \cite{Mizutani1,Mizutani2,Suzuki}. 
We compared 160 nanowire transistors with widths of 3 and 4 nm, utilizing measurements taken on different days. 
The $I_{\rm D}$-$V_{\rm G}$ characteristics of the nanowire transistors are presented in Fig.~\ref{fignanowire}. 
Compared to Fig.1, all the nanowire data appear similar. 
As shown in Fig.~\ref{fignanowire}, 
the $I_{\rm D}$-$V_{\rm G}$ curve of transistor No. 141 (marked by a red circle) is smaller than those of the other transistors.
Here, the device number corresponds to the position of the device in the wafer.

\begin{figure}
\centering
\includegraphics[width=8.5cm]{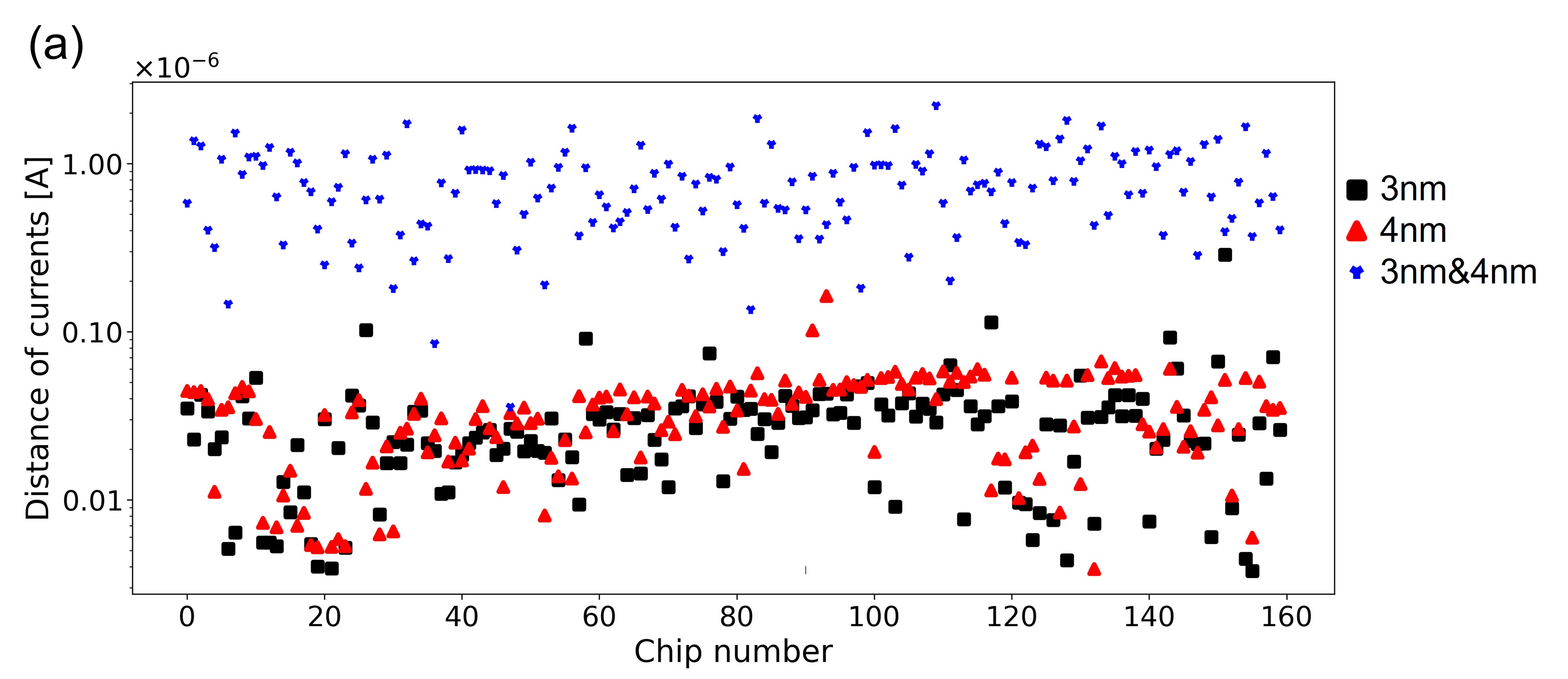}
\includegraphics[width=8.5cm]{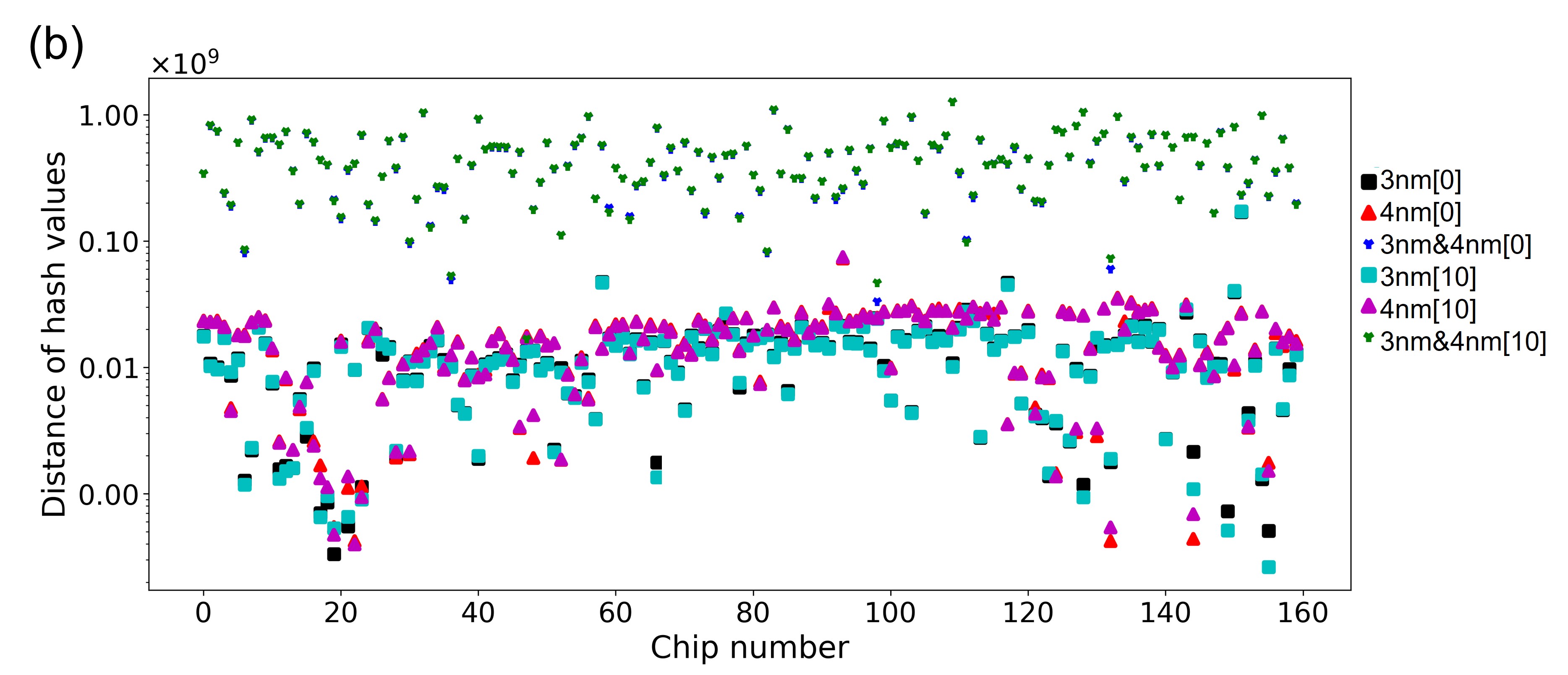}
\caption{
(a) Euclidean distances between the datasets from day 1 and day 2 for 3 and 4 nm nanowire transistors. 
"3nm \& 4nm" represents the Euclidean distance between the datasets of 3 nm-day1 and 4 nm-day1.
(b) Differences in the hash values for the same dataset as in (a) for nanowire transistors. 
The random numbers $\{ {\bf a}, b\}$ are varied, where "[0]" and "[10]" correspond to the results from the first and 11 th sets of random numbers.
}
\label{fighira1}
\end{figure}

First, we estimate the Euclidean distance of the \(I_{\rm D}\)-\(V_{\rm G}\) data as \(n_d\)-dimensional data ($n_d=160$). 
Figure~\ref{fighira1} (a) shows the Euclidean distance between two sets of data observed on different days. 
The same chip measured on different days (day1 and day2) exhibits small distances, 
whereas the distances between different sizes (3 and 4 nm) are considerably larger.
The average Euclidean distances between the data are as follows:
\(d_{3{\rm nm}_{\rm day1}, 3{\rm nm}_{\rm day2}} = 2.983 \times 10^{-8}\) A, 
\(d_{4{\rm nm}_{\rm day1}, 4{\rm nm}_{\rm day2}} = 3.436 \times 10^{-8}\) A, 
\(d_{3{\rm nm}_{\rm day1}, 4{\rm nm}_{\rm day1}} = 7.811 \times 10^{-7}\) A, 
\(d_{3{\rm nm}_{\rm day2}, 4{\rm nm}_{\rm day2}} = 7.879 \times 10^{-7}\) A.
Thus, the Euclidean distance effectively distinguishes devices of different sizes. 
However, as previously mentioned, managing individual \(I\)-\(V\) data becomes increasingly cumbersome as the number of devices grows.

To address this issue, we apply indexing using the hash function. Figure~\ref{fighira1}(b) illustrates the distances between two sets of hash values collected on different days. 
We have examined the hash values by varying the random numbers $\{ {\bf a}, b\}$. 
"[0]" and "[10]" display the results of the first and 11th random number sets.
The parameter \( w = 10^{-9} \) is selected as the standard deviation of the data, 
where a detailed analysis of $w$ will be conducted later.
It can be observed that the distances between devices of different sizes are significantly larger than those between the same devices measured on different days. This indicates that, 
instead of relying on the direct data presented in Fig.~\ref{fighira1}(a), 
the hash values effectively represent the current-voltage characteristics, providing a more efficient approach to distinguishing between devices.

\begin{figure}
\centering
\includegraphics[width=8.5cm]{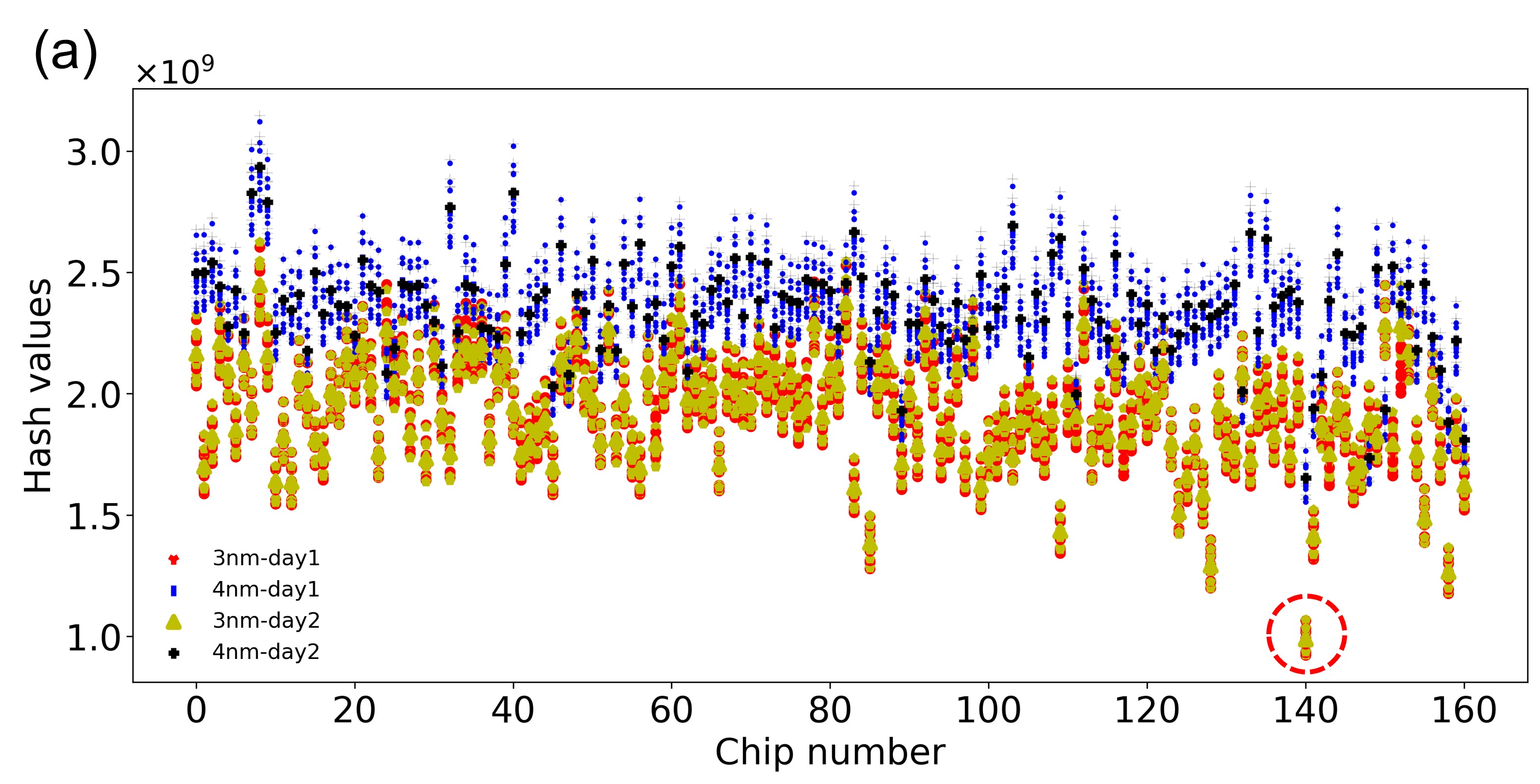}
\includegraphics[width=8.0cm]{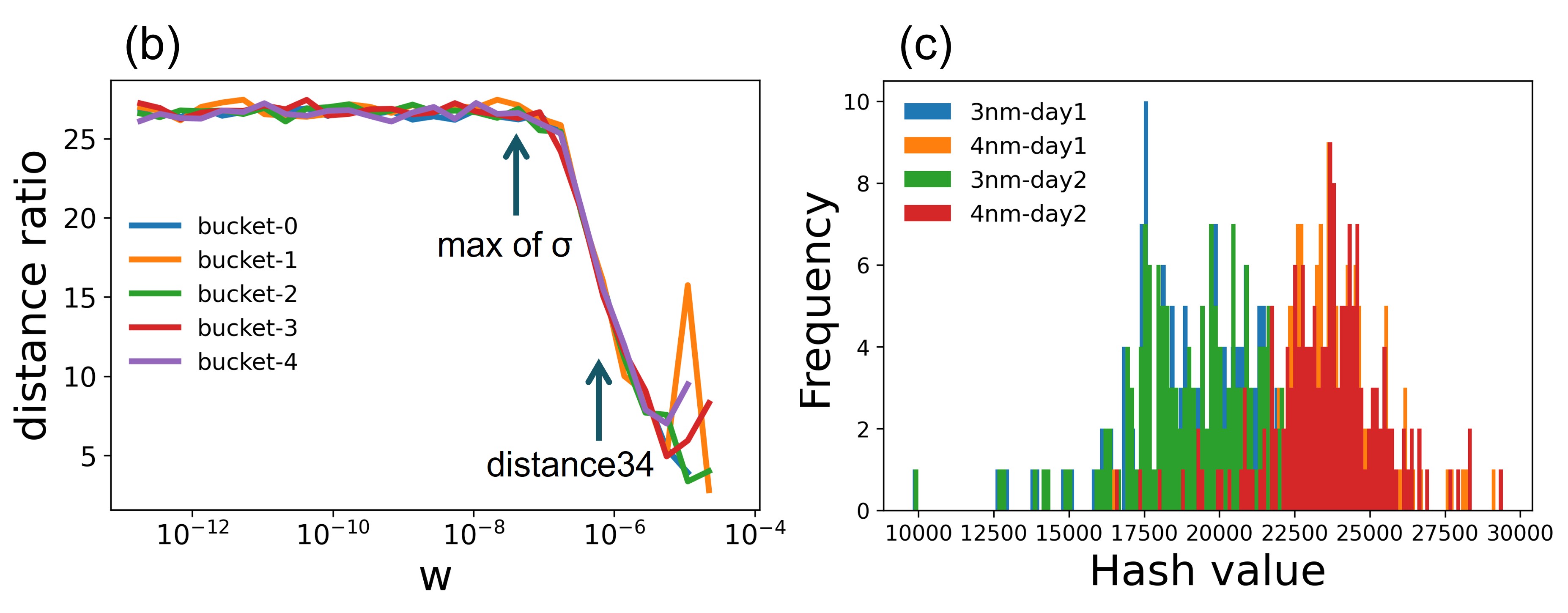}
\caption{
(a) Hash values for the 3 and 4 nm nanowire transistors when the random numbers ${\bf a}$ and $b$ are changed.
varied. The dotted circle highlights chip No. 141, corresponding to Fig.~\ref{fignanowire}.(b)Histogram of hash value distributions.
(c)Distance ratio defined Eq.(3) as the function of $w$.
}
\label{fighira1hash}
\end{figure}

Here, we analyze the suitable range for the hash values. Figure~\ref{fighira1hash}(a) shows the distribution of hash values for four devices, generated using different sets of random numbers, ${\bf a}$ and $b$ (where $N_{\rm buk}=10$). 
It is evident that the hash values shift consistently in the same direction. This observation suggests that while changing the set of random numbers $\{{\bf a}, b\}$ affects the absolute hash values, the relative hash distances between data points remain unchanged.
Focusing on the hash value for chip number 141 (highlighted by the red dotted circles in Figs.\ref{fignanowire} and \ref{fighira1hash}(a)), we observe that its hash value is smaller compared to the other devices. 
This distinction allows us to differentiate devices with varying $I$-$V$ characteristics using their corresponding hash values.
Figure~\ref{fighira1hash}(b) illustrates the dependence of the distance ratio on the parameter $w$, which is defined by:

\begin{equation}
{\rm ratio}_{distance} \equiv 
\frac{dh_{3nm_{\rm day1},4nm_{\rm day1}}+dh_{3nm_{\rm day2},4nm_{\rm day2}}}
{dh_{3nm_{\rm day1},3nm_{\rm day2}}+dh_{4nm_{\rm day1},4nm_{\rm day2}}}, 
\end{equation}
where $dh_{A,B}$ represents the difference between the two hash values of datasets $A$ and $B$.
The "bucket" displays different sets of $\{{\bf a}, b\}$.
In Fig.~\ref{fighira1hash}(b), "$\sigma$" denotes the average standard deviation, 
which is calculated to be 4.56$\times 10^{-8}$ A, 
and "distance34" indicates the average Euclidean distance between the 3 and 4 nm devices. 
A larger distance ratio is beneficial for distinguishing between two devices. 
Therefore, it is preferable to choose $w$ close to the standard deviation, 
which aligns with the findings presented in Fig.~1.

Figure~\ref{fighira1hash}(c) presents the histogram of the hash value distributions, showing that the 3 and 4 nm devices exhibit distinct peak positions, 
which enables differentiation between the two sets of transistors overall. 
However, within the hash value range of 1.6$\times 10^9$ to 2.2$\times 10^9$, 
both 3  and 4 nm devices share similar hash values, making it impossible to determine whether a device belongs to the 3  or 4 nm transistor category in this range. 
This result suggests that relying on a single hash value is insufficient for effectively distinguishing between the two datasets.

\begin{figure}
\includegraphics[width=8.5 cm]{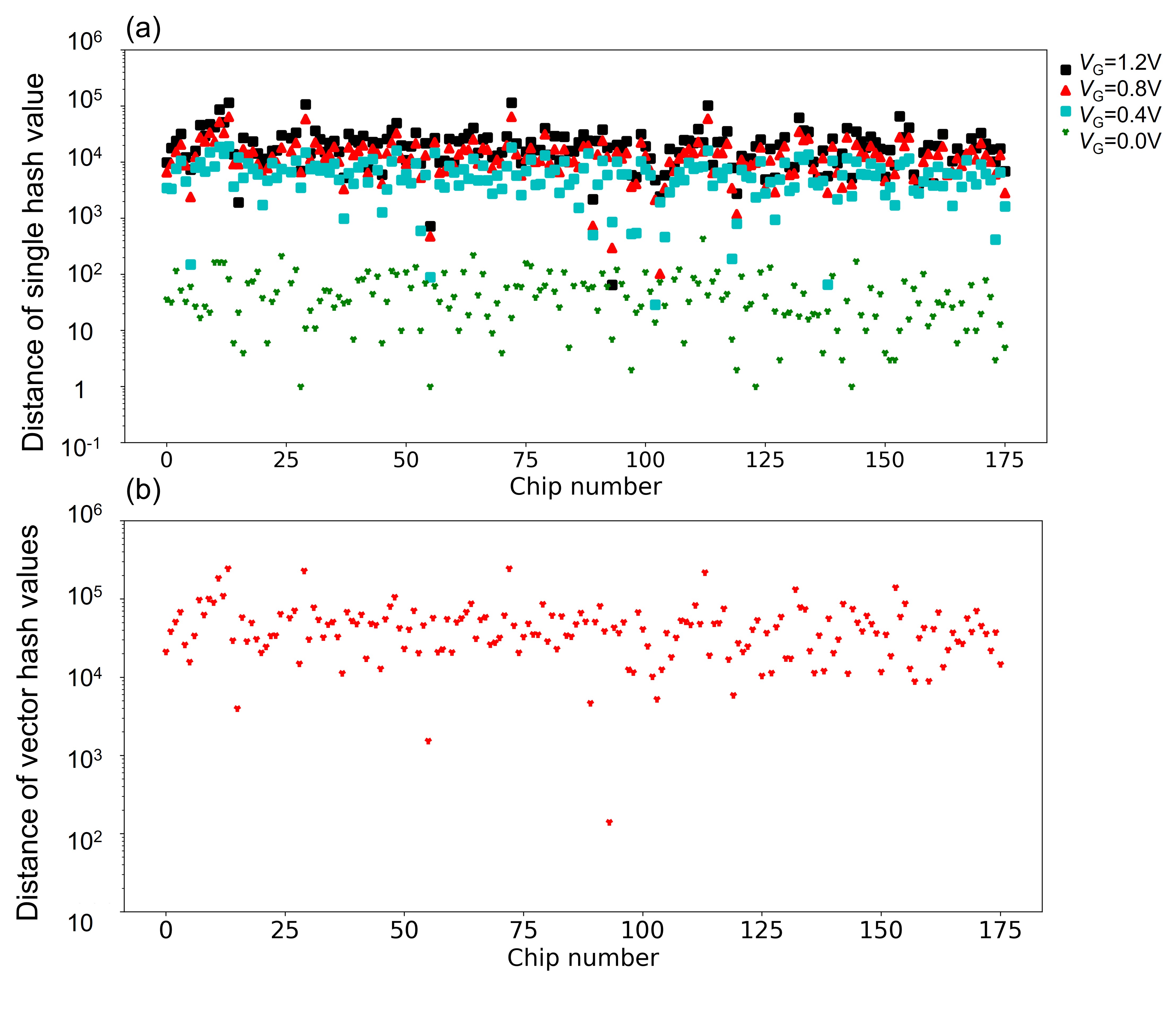}
\caption{
(a) Distance of hash values of $I_{\rm D}$-$V_{\rm D}$ of the 3 and 4 nm nanowires 
for different $V_{\rm G}s$.
(b) Distance of hash values of $I_{\rm D}$-$V_{\rm D}$ of the 3 and 4 nm nanowires
when the data (a) are treated as a vectors.
}
\label{fighira2}
\end{figure}
In conventional transistors, both 
$I_{\rm D}$-$V_{\rm D}$ and $I_{\rm D}$-$V_{\rm G}$ characteristics are provided, allowing us to generate more than a single hash value. This enables the measured values to be treated as vectors. 
Figure~\ref{fighira2}(a) illustrates the difference in hash values between 3 nm and 4 nm nanowires for various $V_{\rm G}$ values. 
Figure~\ref{fighira2}(b) presents the Euclidean distance between the two vectors, 
which are defined as $( h^{3nm}_{V_G=1.2{\rm V}}, h^{3nm}_{V_G=0.8{\rm V}}, h^{3nm}_{V_G=0.4{\rm V}}, h^{3nm}_{V_G=0.0{\rm V}})$ and $( h^{4nm}_{V_G=1.2{\rm V}}, h^{4nm}_{V_G=0.8{\rm V}}, 
h^{4nm}_{V_G=0.4{\rm V}}, h^{4nm}_{V_G=0.0{\rm V}})$. 
Compared to Figure~\ref{fighira2}(a), the minimum distance between the vectors in Fig.~\ref{fighira2}(b) has increased, making the distinction between the two devices more pronounced.

In summary, the application of LSH for device indexing was proposed to differentiate between nearly identical devices that are originally designed to exhibit the same characteristics. 
The LSH algorithm facilitates classification by generating simple hashed numbers, enabling an approximate but effective distinction between these devices.
When multiple types of data, such as $I_{\rm D}$-$V_{\rm D}$ and $I_{\rm D}$-$V_{\rm G}$, are available, 
combining their corresponding hashed identifiers improves classification accuracy. 
The determination of LSH parameters was demonstrated using experimental data from Coulomb blockade phenomena in PMOSFETs and nanowire transistors. 
This classification approach is expected to contribute to device performance enhancement by integrating feedback into the fabrication process.

\subsection*{DATA AVAILABILITY}
The data that supports the findings of this study are available within the article.

\begin{acknowledgments}
This work was supported by JSPS KAKENHI Grant Number JP22K03497.
\end{acknowledgments}


\subsection*{CONFLICT OF INTEREST}
The authors have no conflicts to disclose.

\appendix

\section{clustering approach}
\begin{figure}
\centering
\includegraphics[width=5.5cm]{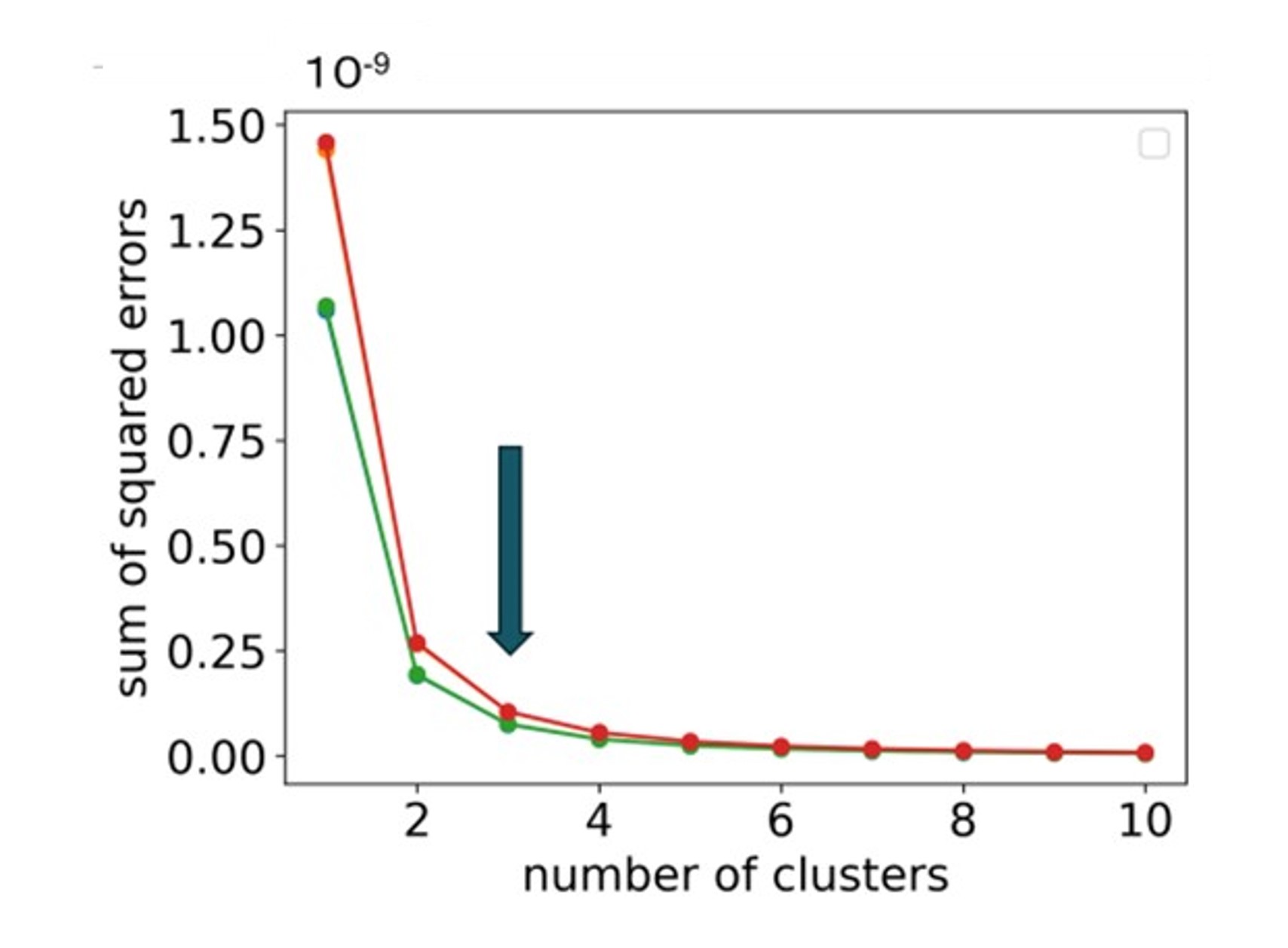}
\caption{
Results of the elbow method to the 
$I_{\rm D}$-$V_{\rm D}$ characteristics of the nanowire transistor.
In the elbow method, the change of the "sum of the square error" represents the plausible number of the clusters.
}
\label{figcluster}
\end{figure}

We perform clustering analysis on silicon nanowire transistors using the Python scikit-learn library~\cite{python}. Clustering algorithms help identify patterns in datasets, 
with K-means being one of the most widely used methods for partitioning data into $k$ clusters.
 To determine an appropriate value for $k$, we apply the elbow method.
Figure~\ref{figcluster} presents the results of the elbow method, suggesting that the $I_{\rm D}$-$V_{\rm D}$ data can be divided into approximately three clusters. This indicates that the dataset consists of similar transistor characteristics.

Next, we apply the k-means algorithm to partition the dataset. When selecting $k=6$, the 168 transistors are categorized into distinct cluster numbers, as detailed in the following analysis.
[
4 5 2 3 4 2 4 1 3 4 5 2 5 4 1 2 2 1 1 4 4 3 4 2 3 4 4 2 4 2 3 1 2 4 3 4 3 
 2 4 4 1 2 2 2 1 5 4 4 3 4 1 2 3 2 1 2 5 4 2 4 4 3 1 4 1 1 2 4 4 1 1 4 4 4
 1 4 1 1 3 1 4 4 3 5 4 0 4 4 1 2 1 2 3 4 2 2 4 5 3 5 2 2 1 2 1 1 2 2 1 0 1
 1 3 2 1 2 4 2 1 4 1 1 4 2 5 5 2 5 0 1 2 2 4 2 1 1 2 4 2 1 \underline{0} 0 1 2 1 1 5 1
 2 3 2 3 4 2 5 1 2 0 2 5]. 
The clustering appears to progress sequentially from transistor No. 1 to No. 160. 
However, the distinct device (transistor No. 141 (underlined)) in Fig.2 (a)(b) is classified into the same cluster as transistors No. 142 and others. 
This instance indicates that the clustering was not entirely successful. 
Therefore, using k-means clustering to distinguish similar data may not always be effective.

\begin{figure}
\centering
\includegraphics[width=5.5cm]{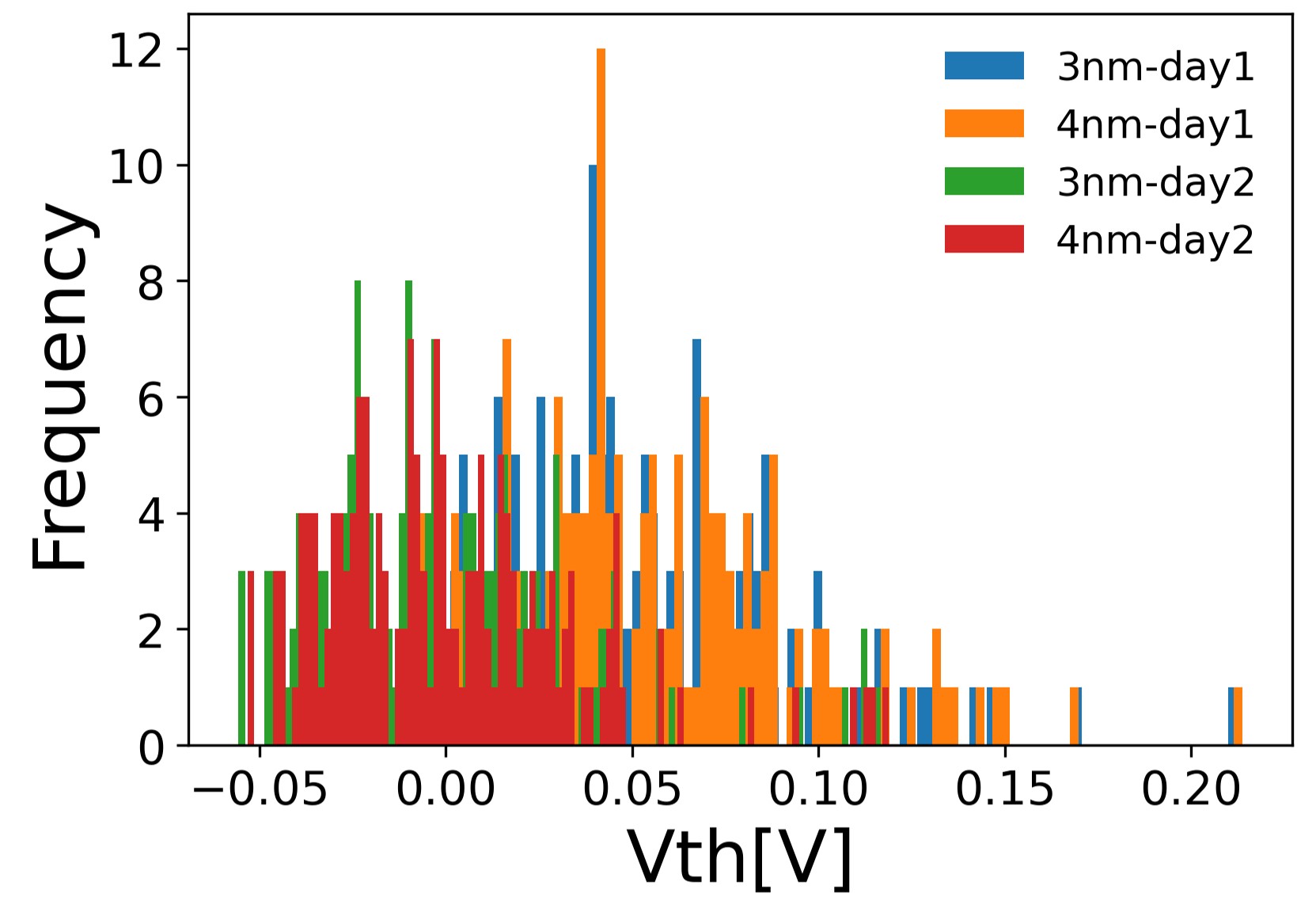}
\caption{
Threshold voltage distributions for the 3nm and 4nm nanowire transistors.
}
\label{figVth}
\end{figure}

Figure~\ref{figVth} displays the histogram of threshold voltages estimated using the conventional method. 
It is evident that the threshold voltages are randomly distributed among the four devices, 
making it impossible to distinguish between them.

\end{document}